\documentclass[12pt,preprint]{aastex}

\newcommand\msol{{\cal M_{\odot}}}

\newcommand\mtr{{{\cal M}_{\rm tr}}}
\newcommand\teff{{T_{\rm eff}}}
\newcommand\amlt{{\alpha_{\rm MLT}}}

\newcommand\lta{\mathrel{\hbox{\raise 0.6 ex \hbox{$<$}\kern
                   -1.8 ex\lower .5 ex\hbox{$\sim$}}}}
\newcommand\gta{\mathrel{\hbox{\raise 0.6 ex \hbox{$>$}\kern
                   -1.7 ex\lower .5 ex\hbox{$\sim$}}}}
 
\shortauthors{VandenBerg et al.}
\shorttitle{Constraint on $Z_\odot$ from M$\,$67}
 
\begin{document}
 
\title{A Constraint on $Z_\odot$ From Fits of Isochrones to the Color-Magnitude
Diagram of M$\,$67}
 
\author{Don A.~VandenBerg\altaffilmark{1}, Bengt Gustafsson\altaffilmark{2},
Bengt Edvardsson\altaffilmark{2}, Kjell Eriksson\altaffilmark{2}, and \\
Jason Ferguson\altaffilmark{3}}

\altaffiltext{1}{Department of Physics \& Astronomy, University of Victoria,
  P.O.~Box 3055, Victoria, B.C., V8W~3P6, Canada (email: vandenbe@uvic.ca)}
\altaffiltext{2}{Uppsala Astronomical Observatory, Box 515, SE-751 20 Uppsala,
Sweden \\ (email: firstname.lastname@astro.uu.se)}
\altaffiltext{3}{Department of Physics, Wichita State University, Wichita,
       Kansas, 67260-0032, U.S.A. \\ (email: jason.ferguson@wichita.edu)}
 
\begin{abstract}
The mass at which a transition is made between stars that have radiative {\it
or} convective cores throughout the core H-burning phase is a fairly sensitive
function of $Z$ (particularly, the CNO abundances).  As a consequence, the
$\sim 4$ Gyr, open cluster M$\,$67 provides a constraint on $Z_\odot$ (and
the solar heavy-element mixture) because (i) high-resolution spectroscopy
indicates that this system has virtually the same metal abundances as the Sun,
and (ii) its turnoff stars have masses just above the lower limit for sustained
core convection on the main sequence.  In this study, evolutionary tracks and
isochrones using the latest MARCS model atmospheres as boundary conditions have
been computed for 0.6--$1.4 \msol$ on the assumption of a metals mix (implying
$Z_\odot \approx 0.0125$) based on the solar abundances derived by M.~Asplund
and collaborators using 3-D model atmospheres.  These calculations do not
predict a turnoff gap where one is observed in M$\,$67.   No such difficulty is
found if the analysis uses isochrones for $Z_\odot = 0.0165$, assuming the
Grevesse \& Sauval (1998) mix of heavy elements.  Our findings, like the
inferences from helioseismology, indicate a problem with the Asplund et 
al.~abundances.  However, it is possible that low-$Z$ models with diffusive
processes taken into account will be less problematic.
\end{abstract} 

 
\section{Introduction}
\label{sec:intro}
The metallicity of the Sun (both the mix of heavy elements, and the total
mass-fraction abundance, $Z$) is presently a subject of considerable
controversy.  Until a few years ago, the generally accepted value of $Z_\odot$
was $\approx 0.017$--0.02, and there was good consistency between the
predictions of Standard Solar Models (SSMs) for metallicities within this range
(see, e.g., \citealt{trm98}, \citealt{bpb01}) and the measured neutrino flux as
well as helioseismological data.  However, significantly reduced abundances for
several of the most abundant heavy elements (including C, N, O, and Ne),
resulting in $Z_\odot \approx 0.0125$, have been derived by M.~Asplund and
collaborators from their analyses of the photospheric spectrum using 3-D,
non-LTE model atmospheres (see \citealt{ags06}; and references therein).
Whereas high-$Z$ solar models are able to reproduce the inferred radial
dependence of the square of the sound speed (down to $\sim 0.1 R_\odot$) from
solar oscillations to $\lta 0.3$\% (see \citealt{cd02}), those computed for the
Asplund et al.~metallicity are unable to do so to better than $\sim 3$\% (e.g.,
(\citealt{tcp04}).

It seems unlikely that this difficulty is due to problems with current opacities
(\citealt{ab05}) or the presently accepted rates of diffusive processes
(\citealt{gu05}).  As noted by \citet{mmt06}, any substantial increases to the
diffusion coefficients that are made in order to recover good agreement with the
inferred sound speed profile result in predictions for the surface helium
content of the Sun that are below the values obtained by the inversion of solar
oscillations.  In fact, turbulent mixing processes appear to operate in stellar
envelopes to {\it reduce} the effects of diffusion on the surface abundances of
$\sim 1 \msol$ main-sequence (MS) stars (see \citealt{kgr06}; \citealt{rmr02}).

While the low solar $Z$ derived by Asplund et al.~may be erroneous, the 3-D
model atmospheres used in their studies have had unprecedented success in
modelling the properties of the solar atmosphere (see \citealt{asp05}), and the
predicted spectral line profiles based on these atmospheres provide superb
matches to the observed profiles (\citealt{ant00}).  In addition, the reduced
abundances of the CNO elements are in excellent agreement with the values
measured in the interstellar medium (\citealt{tcp04}) and in solar neighborhood
B stars (see \citealt{ags05}; and references therein).  Thus, their findings
are not easily dismissed.

There is another, albeit indirect, way of constraining the metallicity of the
Sun that has not yet been investigated.  The large reduction in the CNO and Ne
abundances found by Asplund et al.~(relative to previous estimates) affects the
mass at which a transition is made between stars that have radiative cores 
during the MS phase and those possessing convective cores when their central
hydrogen fuel is exhausted.  Consequently, a cluster like M$\,$67, which has
[m/H] $\approx 0.0$ and turnoff stars with masses close to the transition mass,
$\mtr$, provides a probe of both stellar physics {\it and} the chemical
composition of the material out of which the stars formed.  As shown by
\citet{vbd06}, $\mtr$ is predicted to increase with decreasing metal abundance,
and to be a fairly sensitive function of $Z$ (see their Fig.~2).  Whether or
not an old open cluster is expected to have a gap in its distribution of
near-turnoff stars on the color-magnitude diagram (CMD) will therefore depend
on its metallicity.  (A gap is the observational manifestation of the rapid
contraction phase that accompanies the exhaustion of hydrogen in stars that have
convective cores at the end of their MS lifetimes.)

The M$\,$67 CMD (see \citealt{mmj93}; \citealt{sa04}) contains such a gap,
and stellar models for $Z \gta 0.017$ have been quite successful in reproducing
these data.  \citet{vs04} have shown that a 4.0 Gyr isochrone for $Z = 0.0173$,
assuming the \citet{gn93} metal abundances, provides a satisfactory match to the
cluster CMD, if a small amount of convective core overshooting is assumed (also
see VBD06).  A nearly identical fit was obtained by \citet{mrr04} using
diffusive isochrones for 3.7 Gyr and $Z_{\rm initial} = 0.02$, without having to
postulate any core overshooting.  These calculations predict that the cluster
MS stars have $\sim 6$--12\% surface underabundances of the metals, depending
on their temperatures --- compared with a $\lta 8$\% reduction in the case of
the Sun (\citealt{trm98}), which is much too small to simulate the difference
between the Asplund et al.~solar composition and that reported by, e.g.,
Grevesse \& Noels.

Given the apparent success of relatively metal-rich models in explaining the
CMD of M$\,$67, which has very close to the solar metallicity, is it possible
for $Z=0.0125$ isochrones to provide a comparable fit?  This is the question
that has motivated the present investigation.

\section{Stellar Models for Solar Abundances}
\label{sec:models}
The evolutionary tracks needed for this investigation have been computed using
the Victoria stellar evolution code (see VBD06, and references therein).  Model
atmospheres produced by the latest version of the MARCS code (e.g.,
\citealt{gee03}) have been used as boundary conditions.  [Differences in the
treatment of the surface layers do not alter the tracks for the MS and early
subgiant branch (SGB) phases of solar metallicity stars in any significant way
if the value of the mixing-length parameter, $\amlt$, is chosen so that the
solar constraint is satisfied (see, e.g., Fig.~3 in the study by
\citealt{vee07}).  The assumed value of $\amlt$ does affect the predicted
location of the red-giant branch (RGB), but that is not a concern for the
present work.]

The standard abundance distribution reported by \citet{gs98}, as determined from
analyses of solar photospheric spectra and meteoritic data using classical 1-D
hydrostatic models, was the last such compilation prior to the use of 3-D
hydrodynamical model atmospheres by Asplund et al.~(2006) to determine the solar
composition.  The GS98 abundances for the 19 metals normally considered in
stellar evolutionary computations (because they are the only heavy elements that
are taken into account in the calculation of OPAL opacities for stellar interior
conditions; see \citealt{ir96}) are listed in the second column of
Table~\ref{tab:tab1}.  The third column contains the the abundance distribution
that is obtained when the revised values derived by Asplund et al.~for several
of the heavy elements are adopted: note that the C and N abundances in
Table~\ref{tab:tab1} are 0.02 dex higher than their published 2006 estimates.

OPAL opacities were computed for both heavy-element mixtures
using the Livermore Laboratory website
facility (see http://www-phys.llnl.gov/Research/OPAL).  Because these
calculations do not include the contributions from molecular sources or grains,
complementary opacity data for $T \le 10^4$ K were computed for the same
abundances using the code described by \citet{faa05}.  The tabulated values of
$\log N_{\rm He}$ were obtained from SSMs that were constructed for the two
sets of abundances: the resultant values of $Y$ and $Z$ are listed in the last
two rows of Table~\ref{tab:tab1}.  To satisfy the solar constraint, the
$1.0 \msol$ models for the GS98 and Asplund et al.~abundances required
$\amlt = 1.84$ and 1.80, respectively (when MARCS model atmospheres are used as
boundary conditions; see \citealt{vee07}).

The value of $\mtr$ is a function of $Z$ (especially the CNO abundances; see
below).  The transition occurs when the CNO-cycle becomes an important source of
nuclear energy production, and since both the decrease in the {\it abundances}
of the CNO elements and the concomitant reduction in opacity will serve to
reduce the rate of the CNO-cycle, the minimum mass for sustained core convection
on the MS is higher in stars of lower $Z$.  By trial and error, $\mtr$ was found
to be $1.195 \msol$ if $Z = 0.0125$ (Asplund et al.~metals mix), whereas $1.155
\msol$ is obtained if $Z = 0.0165$ (GS98 metal abundances).  [The transition
mass also depends on the adopted helium content; e.g., $\mtr = 1.181 \msol$ if
$Z=0.0125$ and the value of $Y$ required by an SSM for $Z=0.0165$ is assumed 
(i.e., $Y=0.2676$; see Table~\ref{tab:tab1}).]

\begin{figure}
\plotone{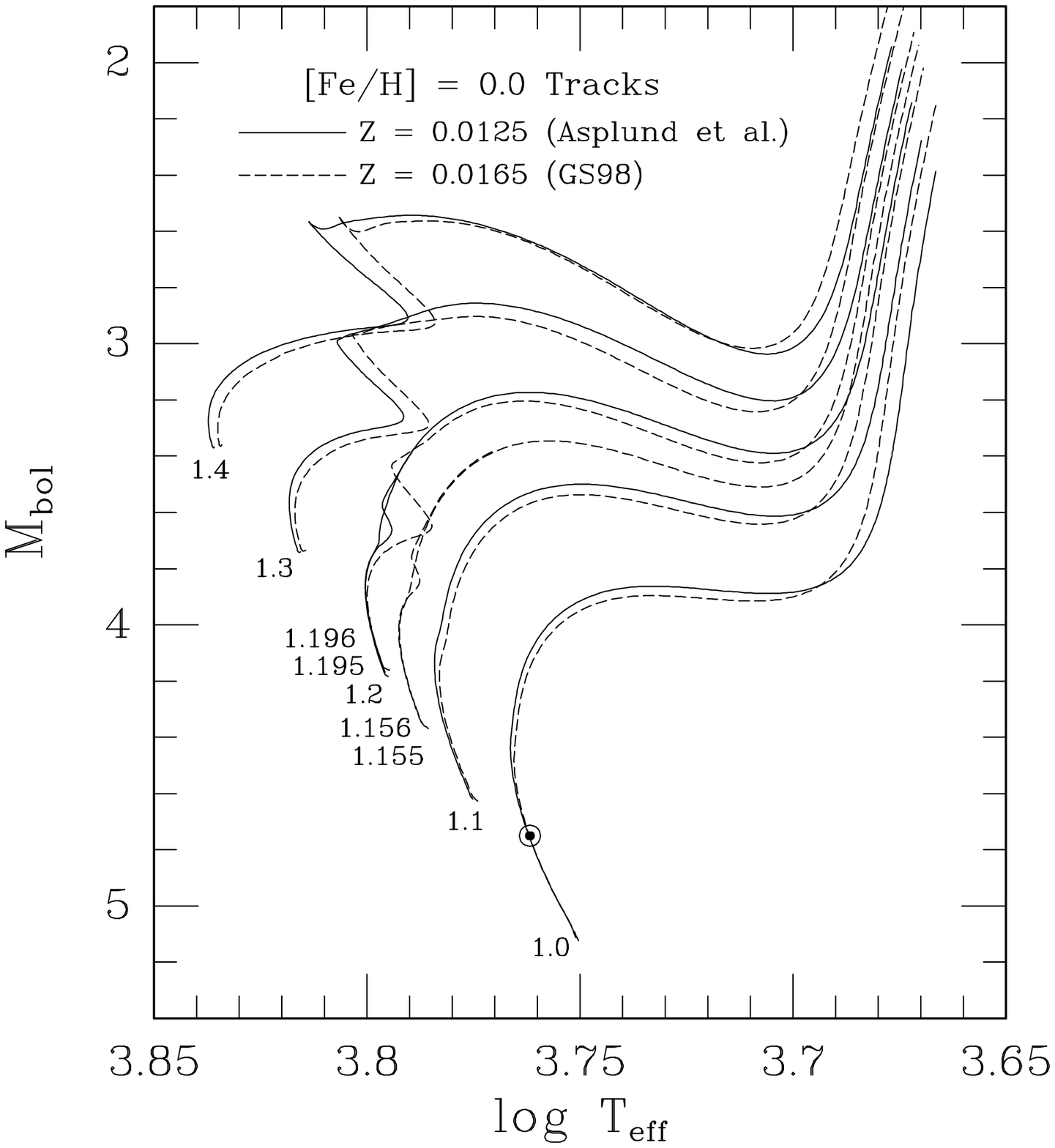}
\caption{Grids of evolutionary tracks for the indicated values of $Z$ and masses
(in solar units).  The sequences for 1.155 and $1.156 \msol$ (dashed curves) and
for 1.195 and $1.196 \msol$ (solid curves) show how rapidly the morphology of
the tracks changes near $\mtr$ (see the text). Although not plotted, both grids
were extended down to $0.6 \msol$.}
\label{fig:vfig1}
\end{figure}

Indeed, the character of the tracks changes abruptly when core convection
persists until H exhaustion.  This is shown in Figure~\ref{fig:vfig1}, which
plots the evolutionary tracks for 1.0--$1.4 \msol$ that were computed for the
two values of $Z_\odot$; including, in particular, sequences for $\mtr$ and
$(\mtr + 0.001) \msol$.  Each of the higher mass tracks possesses a blueward
hook, which arises when a star contracts following the depletion of hydrogen in
a convective core in order to ignite H-burning in a shell around the He core.
This morphology is not seen in the lower mass tracks because the central regions
are radiative when hydrogen is exhausted and there is a smooth transition to 
H shell burning.

A difference of $0.04 \msol$ in $\mtr$ is surprisingly large given that VBD06
found an increase of only $0.009 \msol$ when $Z$ was decreased from 0.0173 to
0.0125.  However, the latter calculations assumed the same relative abundances
of the metals (those given by \citealt{gn93}), whereas the tracks plotted in
Fig.~\ref{fig:vfig1} assume two very different heavy-element mixtures.  Indeed,
a large reduction in the CNO abundances is mainly responsible for the downward
revision of the solar value of $Z$ from 0.0165 (GS98) to 0.0125 (Asplund et
al.) --- which demonstrates that $\mtr$ is a strong function of the abundances
of the CNO elements.

An increase in $\mtr$ from 1.155 to $1.195 \msol$ will be accompanied by a
decrease in the maximum age at which a gap near the turnoff is expected in an
observed CMD.  Fig.~\ref{fig:vfig1} shows that the biggest differences between
the two sets of models occur at masses of $\approx 1.2 \msol$.  Because $\mtr$
is lower in the case of the GS98 computations, and because the amount of
overshooting in stars of mass ${\cal M_*} > \mtr$ is assumed to vary directly
with ${\cal M_*} - \mtr$ (see VBD06), the tracks represented by the dashed
curves for masses of 1.2--$1.4 \msol$ have larger convective cores during MS
evolution, and cooler blueward hooks at H exhaustion, than those plotted as
solid curves.  Such differences will impact the isochrones (computed in this 
study using the interpolation software described by VBD06) and the degree to
which they are able to reproduce the M$\,$67 CMD.

\section{Application to the M$\,$67 CMD}
\label{sec:m67}

According to high-resolution spectroscopy, M$\,$67 has [Fe/H] $= 0.0 \pm 0.03$
(see \citealt{tet00}; \citealt{rsp06}).  For the cluster reddening, we have
adopted $E(B-V) = 0.038$ mag, as found from the \citet{sfd98} dust maps: this
estimate is probably accurate to within $\pm 0.01$ mag, given that it agrees so
well with the values derived using alternative methods (see \citealt{ntc87};
\citealt{svk99}).  As far as the distance is concerned, the modulus that is
obtained from a main-sequence fit of the dereddened photometry to the lower MS
segments of our isochrones should also be quite accurate (probably to within
$\pm 0.1$ mag) because the model temperatures satisfy the solar constraint and
the adopted color transformations (by \citealt{vc03}) are in very good agreement
with the empirical relations derived by \citet{sf00} from their study of solar
neighborhood stars having well-determined properties (see the VandenBerg \& Clem
paper).  The application of the MS-fitting technique yields $(m-M)_V = 9.70$.

\begin{figure}
\plotone{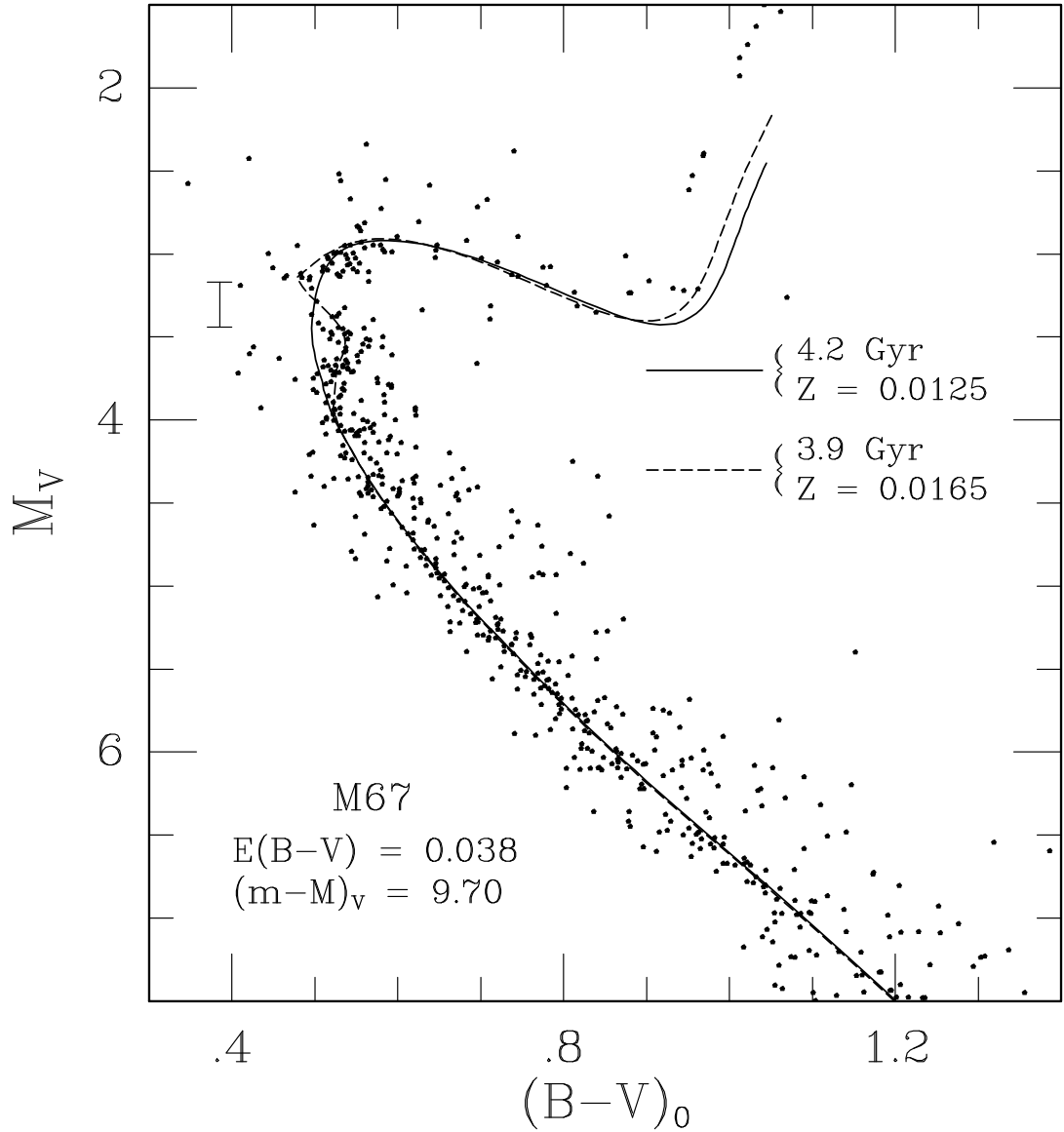}
\caption{Comparison of isochrones for the indicated and ages and metallicities
with the CMD of M$\,$67 (Montgomery et al.~1993), on the assumption of the
reddening and apparent distance modulus specified in the lower left-hand
corner.  The vertical line bounded by short horizontal lines, located adjacent
to the turnoff, indicates our estimate of the luminosity spanned by the gap
in the observed CMD.  Only the dashed isochrone has a blueward hook in the
same magnitude range.}
\label{fig:vfig2}
\end{figure}

With these choices for the basic parameters of M$\,$67, the isochrone for each
value of $Z_\odot$ that provides the best fit to the subgiants is readily
identified.  As shown in Figure~\ref{fig:vfig2}, the 3.9 Gyr isochrone for
$Z=0.0165$ (GS98 metal abundances) provides a good match to the morphology of
the M$\,$67 CMD all the way from the ZAMS to the base of the giant branch,
including, in particular, the location of the gap near the turnoff.  (Many of
the stars just brighter than the gap, at $M_V \approx 3.1$, are presumably
binaries, given that M$\,$67 is known to have an unusually large binarity
fraction; at least 63\%, according to Montgomery et al.~1993.  The $\sim 0.06$
mag offset in color at a given $M_V$ along the lower RGB is discussed by
VandenBerg et al.~2007.)  In stark contrast with this, the 4.2 Gyr isochrone
for $Z=0.0125$ (the Asplund et  al.~metallicity) fails to reproduce the location
of the near-turnoff stars of M$\,$67, despite providing a comparably good fit
to the ZAMS and the SGB.  This isochrone shows no indication at all of a
blueward hook feature, and thus it does not predict a gap in the distribution
of stars near the turnoff where one is observed.\footnote{Although one might
question the reality of the gap in the Montgomery et al.~(1993) CMD, there is
no doubt about its existence (in the same magnitude range as that indicated in
Fig.~\ref{fig:vfig2}) in the CMD published by \citet{sa04} for the
high-probability single-star members.  Moreover, the observed distribution of
single stars along the cluster fiducial sequence (see Sandquist's Fig.~14) is
very reminiscent of that predicted by best-fitting isochrones (see Fig.~14 in
the study by \citealt{mrr04}).  [The CMD obtained by Sandquist is not used here
because he did not provide $BV$ photometry (only $VI$), and his $V-I$ colors,
but not his $V$ magnitudes, are at odds with other measurements (see the
discussion by \citealt{vs04}).  Perhaps the main advantage of using $BV$ data
in our analysis is that, as mentioned above, the $(B-V)$--$\teff$ relations
that have been used to transpose the models to the observed plane appear to be
especially reliable.]}

The oldest isochrone in the $Z=0.0125$ set that has the required turnoff
morphology is one for an age of 3.6 Gyr, and indeed, it provides a satisfactory
fit to the CMD of M$\,$67 if $E(B-V) \approx 0.065$ and $(m-M)_V \approx 9.90$.
However, such a high reddening seems to be ruled out and it is doubtful that
the model colors are in error by as much as 0.03 mag.  Isochrones for the
Asplund metallicity thus appear to pose similar difficulties for our
understanding of M$\,$67 as for helioseismology, in the sense that {\it stellar
models with higher values of $Z_\odot$ are better able to explain the
observations}.

As the CNO elements play such a key role in the interpretation of the cluster
data, one must first consider whether their abundances, relative to iron, are
higher in M$\,$67 than in the Sun.  An increase in the abundances of C, N,
{\it and} O by $\sim 0.1$ dex (or more, if the increase is limited to one or
two of these elements) may be enough to reconcile the predicted and observed
CMDs.  However, while a true cosmic spread among solar metallicity disk stars
with an amplitude of about $0.1$ dex in [C/Fe], [N/Fe], and [O/Fe] cannot be
excluded (see the recent surveys by \citealt{rtl03} and \citealt{eis06}), the
abundance analysis of M$\,$67 main-sequence stars carried out by \citet{rsp06}
does not support the possibility that M$\,$67 has [CNO/Fe] $\gta 0.1$.  These
authors derived a mean oxygen abundance corresponding to [O/Fe] $= -0.01\pm
0.03$, which agrees well with the findings of \citet{tet00} for cluster giants.
The latter also derived C and N abundances, though, in the case of giant stars,
they are expected to have been altered by CN burning and the first dredge-up.
They find that [C/H] and [N/H] are typically $-0.2$ and $+0.2$, respectively,
which is roughly consistent with initial C and N abundances slightly {\it lower}
than solar.  In fact, \citet{fb92} found a mean [C/H] of $-0.09 \pm 0.03$ for
three F dwarfs of M$\,$67.  [Based on the work of \citet{npa02}, we find that
the expected differential effects due to the use of 3-D model atmospheres in
the determination of [O/Fe] in the cluster turnoff stars, like those analyzed
by Randich et al., would amount to $+0.03$ dex relative to the solar value.]
Thus, it seems improbable that the CNO abundances relative to iron are enhanced
in the cluster with respect to the Sun by as much as $0.1$ dex.  We also note
that an increase of the Ne abundance, which has been suggested (by
\citealt{bbs05}) to solve the discrepancy between solar oscillation data and
the Asplund et al.~abundances would not resolve the M$\,$67 conundrum.


Although our present understanding of diffusive processes precludes the 
possibility that there is a large difference between the interior and 
photospheric metallicities of the Sun, \citet{mrr04} have found from their
calculations for $Z_\odot \ge 0.0175$ that the lowest mass diffusive model with
a convective core on the MS is $\approx 1.10 \msol$, as opposed to $1.14 \msol$
in the case of non-diffusive models.  The difference in mass is comparable to
the difference in $\mtr$ predicted by our calculations when the assumed value
of $Z$ is increased from 0.0125 (Asplund et al.) to 0.0165 (GS98). Consequently,
it seems quite possible that low-$Z$ {\it diffusive} models (with or without
some core overshooting) may be able to satisfy the M$\,$67 constraint.  Further
investigation is needed.  However, even if this leads to a consistent
explanation for both the Asplund et al.~solar abundances and the M$\,$67 CMD,
the large discrepancies with helioseismology would remain.


\section{Conclusions}
\label{sec:conclude}
The new Asplund et al.~metallicity for the Sun presents some difficulties for
fits of solar abundance models to the M$\,$67 CMD, in that they do not predict
a gap near the turnoff where one is observed.  Whether or not stellar models
predict a gap at the observed $M_V$ in this open cluster depends quite
sensitively on the assumed CNO abundances, and it is mainly the revision of
these elemental abundances (along with that of Ne) that is responsible for the
decrease in $Z_\odot$ from 0.0165 (GS98) to 0.0125.  Isochrones for the higher
value of $Z$ are able to reproduce the detailed CMD morphology of M$\,$67 in
the vicinity of the turnoff without apparent problems.  Interestingly, it is
primarily the reduction in the abundances of CNO that has resulted in
substantial difficulties for helioseismology; e.g., an increase from 
$\sim 0.3$\% to $\sim 3$\% in the differences between the predicted and inferred
sound speed squared profiles.  Are solar oscillation studies and our 
investigation of M$\,$67 telling us that the low solar $Z$ determined by
Asplund et al.~is wrong?

Not necessarily.  It is possible, judging from the work of \citet{mrr04},
that models which take diffusive processes into account may not have the same
difficulties as the models used in this study, which neglect such processes.
That is, if the Asplund et al.~solar abundances are correct, {\it only} those
low-$Z$ models that treat diffusion may be able to reproduce the M$\,$67 CMD.
This possibility, which would {\it not} resolve the quandary presented by
solar oscillations, needs to be carefully studied.  Diffusion clearly adds
another level of complexity to the problem since, e.g., the initial abundances
of the gas out of which the Sun and M$\,$67 formed would have been somewhat
different, if they have the same abundances today, because the Sun is up to 1
Gyr older than M$\,$67.  Importantly, MS stars in M$\,$67 should show systematic
variations in their surface metallicities as a function of $\teff$ due to the
operation of diffusive processes (see Michaud et al.).

\acknowledgements
We thank Martin Asplund and Santi Cassisi for valuable suggestions and comments
on the manuscript. This work has been supported by the Natural Sciences and
Engineering Research Council of Canada through a Discovery Grant to DAV, and by
grants from the Swedish Research Council to BE, KE, and BG.  JF acknowledges
support from NSF grant AST-0239590.

\clearpage

\clearpage

\clearpage
\begin{deluxetable}{ccc}
\tabletypesize{\footnotesize}
\tablecaption{Solar Elemental Abundances \label{tab:tab1}}
\tablewidth{0pt}
\tablehead{ & \multispan2\hfil $\log N$\hfil \\ & \multispan2\hrulefill \\ 
Element & Grevesse \& Sauval (1998) & Asplund et al.\tablenotemark{a} } 
\startdata
H\phantom{e} & 12.00\phantom{38}           & 12.00\phantom{87}           \\
He\tablenotemark{b}   & 10.9738            & 10.9487                     \\
C\phantom{e} & \phantom{1}8.52\phantom{38} & \phantom{1}8.41\phantom{87} \\
N\phantom{e} & \phantom{1}7.92\phantom{38} & \phantom{1}7.80\phantom{87} \\
O\phantom{e} & \phantom{1}8.83\phantom{38} & \phantom{1}8.66\phantom{87} \\
Ne           & \phantom{1}8.08\phantom{38} & \phantom{1}7.84\phantom{87} \\
Na           & \phantom{1}6.33\phantom{38} & \phantom{1}6.33\phantom{87} \\
Mg           & \phantom{1}7.58\phantom{38} & \phantom{1}7.58\phantom{87} \\
Al           & \phantom{1}6.47\phantom{38} & \phantom{1}6.47\phantom{87} \\
Si           & \phantom{1}7.55\phantom{38} & \phantom{1}7.51\phantom{87} \\
P\phantom{e} & \phantom{1}5.45\phantom{38} & \phantom{1}5.45\phantom{87} \\
S\phantom{e} & \phantom{1}7.33\phantom{38} & \phantom{1}7.33\phantom{87} \\
Cl           & \phantom{1}5.50\phantom{38} & \phantom{1}5.50\phantom{87} \\
Ar           & \phantom{1}6.40\phantom{38} & \phantom{1}6.18\phantom{87} \\
K\phantom{e} & \phantom{1}5.12\phantom{38} & \phantom{1}5.12\phantom{87} \\
Ca           & \phantom{1}6.36\phantom{38} & \phantom{1}6.36\phantom{87} \\
Ti           & \phantom{1}5.02\phantom{38} & \phantom{1}5.02\phantom{87} \\
Cr           & \phantom{1}5.67\phantom{38} & \phantom{1}5.67\phantom{87} \\
Mn           & \phantom{1}5.39\phantom{38} & \phantom{1}5.39\phantom{87} \\
Fe           & \phantom{1}7.50\phantom{38} & \phantom{1}7.45\phantom{87} \\
Ni           & \phantom{1}6.25\phantom{38} & \phantom{1}6.25\phantom{87} \\
\multispan3\hrulefill \\
$Y$          & 0.2676         & 0.2559 \\
$Z$          & 0.0165         & 0.0125 \\
\enddata
\tablenotetext{a}{Abundances for C, N, O, Ne, Si, Ar, and Fe were provided
by M.~Asplund (2004, priv.~comm.): Grevesse \& Sauval (1998) abundances are
assumed for all other elements heavier than helium.}
\tablenotetext{b}{Determined from Standard Solar Models.}
\end{deluxetable}

\end{document}